# ENERGY-EFFICIENT RESOURCE ALLOCATION IN MULTIUSER MIMO SYSTEMS: A GAME-THEORETIC FRAMEWORK

*Stefano Buzzi*[1], *H. Vincent Poor*[2], *and Daniela Saturnino*[1]

[1]University of Cassino, DAEIMI

03043 Cassino (FR) - Italy; {buzzi, d.saturnino}@unicas.it

[2]Princeton University, School of Engineering and Applied Science

Princeton, NJ, 08544 - USA; poor@princeton.edu

**ABSTRACT**

This paper focuses on the cross-layer issue of resource allocation for energy efficiency in the uplink of a multiuser MIMO wireless communication system. Assuming that all of the transmitters and the uplink receiver are equipped with multiple antennas, the situation considered is that in which each terminal is allowed to vary its transmit power, beamforming vector, and uplink receiver in order to maximize its own utility, which is defined as the ratio of data throughput to transmit power; the case in which non-linear interference cancellation is used at the receiver is also investigated. Applying a game-theoretic formulation, several non-cooperative games for utility maximization are thus formulated, and their performance is compared in terms of achieved average utility, achieved average SINR and average transmit power at the Nash equilibrium. Numerical results show that the use of the proposed cross-layer resource allocation policies brings remarkable advantages to the network performance.

## 1. INTRODUCTION

The increasing demand for new wireless applications, and the tremendous progress in the development of smartphones and handheld devices with exceptional computing capabilities requires wireless communication infrastructures capable of delivering data at higher and higher data-rates. The use of multiple antennas at both ends of a wireless link has proved to be a key technology to improve the spectral efficiency of wireless networks [1]. Likewise, intelligent resource allocation procedures also will play a prominent role to ensure reliability and efficiency in future wireless data networks.

This paper focuses on the uplink of a multiuser multiple-input multiple-output (MIMO) communication system, wherein both the mobile terminals and the common access point (AP) are equipped with multiple antennas. We are interested in the design of non-cooperative resource allocation policies aimed at energy efficiency maximization, which is defined here as the number of reliably delivered information symbols per unit-energy taken from the battery. Energy-efficiency maximization is indeed a crucial problem in mobile wireless communications, wherein mobile users are interested in making a careful and smart use of the energy stored in their battery. Following a recent trend, we use game theory tools [2] in order to obtain non-cooperative resource allocation procedures, maximizing each user's energy efficiency with respect to its own transmit power, beamforming vector and uplink receiver.

A game-theoretic framework for non-cooperative energy efficiency maximization has been widely applied in the recent past to design resource allocation policies for code division multiple access (CDMA) systems [3, 4, 5] and for ultra-wideband (UWB) systems [6]. On the other hand, MIMO communication systems have received a great deal of attention in the last decade (see, for instance, the references in the recent textbook [1]). Among the studies addressing joint transmitter and receiver adaptation for improved performance, we cite the papers [7, 8], which consider transceiver optimization for multiuser MIMO systems in cooperative environments, i.e. assuming that a central processor allocates resources among active users, and neglecting the issue of power control.

In this paper, we extend the game-theoretic framework, surveyed in [9], to multiuser MIMO wireless systems. We consider the case in which energy efficiency is to be maximized with respect to

**a.** the transmit power of each user, assuming matched filtering at the receiver;

**b.** the transmit power and the choice of the uplink linear receiver for each user;

**c.** the transmit power, the beamforming vector and the choice of the uplink linear receiver for each user; and

**d.** the transmit power and the choice of the non-linear serial interference cancellation (SIC) uplink receiver for each user.

Note that consideration of these games is not a trivial extension of the results reported in authors' previous studies, since the analysis of the Nash equilibrium (NE) points for some of the above games, and in particular for the cases **c.** and **d.** poses new mathematical challenges. More precisely, we will see that problems **a.** and **b.** are somewhat equivalent to those treated in [3] and [4] for a CDMA system, while, instead proof of the existence of a NE for problem **c.** requires a new and different approach. Finally, the consideration of problem **d.**, which assumes the use of a non-linear SIC receiver, has not yet appeared in the open literature.

Results will show that the use of advanced resource allocation policies brings remarkable improvements in terms of achieved energy efficiency at the equilibrium thus enabling the transmission of larger bulks of data for a given amount of energy stored in the battery.

This research was supported in part by the U. S. National Science Foundation under Grants ANI-03-38807 and CNS-06-25637.

## 2. PRELIMINARIES AND PROBLEM FORMULATION

Consider the uplink of a $K$-user synchronous, single-cell, MIMO multiuser system subject to flat fading. Denote by $N_T$ the number of transmit antennas for each user, and by $N_R$ the number of receive antennas at the common AP. Collecting in an $N_R$-dimensional vector, say $\mathbf{r}$, the samples at the output of the receiver front-end filter and corresponding to one symbol interval, we have

$$\mathbf{r} = \sum_{k=1}^{K} \sqrt{p_k} \mathbf{H}_k \mathbf{a}_k b_k + \mathbf{n} , \qquad (1)$$

wherein $p_k$ is the transmit power of the $k$-th user[1], $b_k \in \{-1,1\}$ is the information symbol of the $k$-th user, and $\mathbf{H}_k$ is the real[2] $N_R \times N_T$ matrix channel gain between the $k$-th user's transmitter and the AP; the entries of $\mathbf{H}_k$ depend on both the distance of the $k$-th user's terminal from the AP and on the fading fluctuations. The $N_T$-dimensional vector $\mathbf{a}_k$ is the beamforming vector of the $k$-th user; we assume that $\mathbf{a}_k^T \mathbf{a}_k = \|\mathbf{a}_k\|^2 = 1$, with $(\cdot)^T$ denoting transpose. Finally, $\mathbf{n}$ is the ambient noise vector, which we assume to be a zero-mean white Gaussian random process with covariance matrix $(\mathcal{N}_0/2)\mathbf{I}_{N_R}$, with $\mathbf{I}_{N_R}$ the identity matrix of order $N_R$.

Assume now that each mobile terminal sends its data in packets of $M$ bits, and that it is interested both in having its data received with as small as possible error probability at the AP, and in making careful use of the energy stored in its battery. Obviously, these are conflicting goals, since error-free reception may be achieved by increasing the transmit power, which of course comes at the expense of battery life. A useful approach to quantify these conflicting goals is to define the utility of the $k$-th user as the ratio of its throughput, defined as the number of information bits that are received with no error in unit time, to its transmit power [3] - [6], i.e.

$$u_k = T_k / p_k . \qquad (2)$$

Note that $u_k$ is measured in bit/Joule, i.e. it represents the number of successful bit transmissions that can be made for each energy-unit drained from the battery.

Denoting by $R$ the common rate of the network (extension to the case in which each user transmits with its own rate $R_k$ is quite simple) and assuming that each packet of $M$ symbols contains $L$ information symbols and $M-L$ overhead symbols, reserved, e.g., for channel estimation and/or parity checks, denoting by $\gamma_k$ the Signal-to-Interference-plus-Noise Ratio (SINR) for the $k$-th user at the receiver output, and following the reasoning of [3, 4], a faithful and mathematically tractable approximation for the utility $u_k$ in (2) is the following

$$u_k = R \frac{L}{M} \frac{f(\gamma_k)}{p_k} , \quad \forall k = 1, \ldots, K . \qquad (3)$$

In the above equation, $f(\gamma_k)$ is the so-called *efficiency function*, approximating the probability of successful (i.e. error-free) packet reception. As an example, for BPSK modulation, the choice $f(\gamma_k) = (1 - e^{-\gamma_k})^M$ is widely accepted. The results of this paper, however, hold not only for this particular choice, but for any efficiency function $f(\cdot)$ that is increasing, S-shaped, approaching unity as $\gamma_k \to +\infty$, and such that $f(\gamma_k) = o(\gamma_k)$ for vanishing $\gamma_k$.

Now, based on the utility definition (3), many interesting questions arise concerning how each user may maximize its utility, and how this maximization affects utilities achieved by other users. Game theory provides means to study these interactions and to provide some useful and insightful answers to these questions. It has been applied in this context mainly as a tool to study non-cooperative resource allocation procedures for CDMA systems and for UWB communications. In the following, instead, we address the problem of non-cooperative energy efficiency maximization in multiuser MIMO systems.

## 3. NON-COOPERATIVE RESOURCE ALLOCATION: LINEAR RECEIVER

A linear receiver detects the data symbol $b_k$, according to the decision rule

$$\widehat{b}_k = \text{sign} \left[ \mathbf{d}_k^T \mathbf{r} \right] , \qquad (4)$$

with $\widehat{b}_k$ the estimate of $b_k$ and $\mathbf{d}_k \in \mathscr{R}^{N_R}$ the $N_R$-dimensional vector representing the receive filter for the user $k$ (the set $\mathscr{R}$ is the real field). It is easily seen that the SINR $\gamma_k$ can be written as

$$\gamma_k = \frac{p_k (\mathbf{d}_k^T \mathbf{H}_k \mathbf{a}_k)^2}{\frac{\mathcal{N}_0}{2} \|\mathbf{d}_k\|^2 + \sum_{i \neq k} p_i (\mathbf{d}_k^T \mathbf{H}_i \mathbf{a}_i)^2} . \qquad (5)$$

In the following, we consider non-cooperative resource allocation games aimed at energy efficiency maximization with respect to (a) the transmit power, assuming that a matched filter is used at the receiver; (b) the transmit power and the linear uplink receiver; and (c) the transmit power, the beamforming vector and the choice of the linear uplink receiver.

### 3.1 Transmit power control with matched filtering

Assume that the $k$-th user's beamforming vector is taken parallel to the eigenvector corresponding to the maximum eigenvalue of the matrix $\mathbf{H}_k^T \mathbf{H}_k$, and that a classical matched filter is used at the AP, i.e. we have $\mathbf{d}_k = \mathbf{H}_k \mathbf{a}_k$. The $k$-th user SINR is now expressed as

$$\gamma_k = \frac{p_k \|\mathbf{H}_k \mathbf{a}_k\|^4}{\frac{\mathcal{N}_0}{2} \|\mathbf{H}_k \mathbf{a}_k\|^2 + \sum_{i \neq k} p_i (\mathbf{a}_k^T \mathbf{H}_k^T \mathbf{H}_i \mathbf{a}_i)^2} . \qquad (6)$$

Considering the non-cooperative game

$$\max_{p_k \in [0, P_{k,\max}]} \frac{f(\gamma_k)}{p_k} , \quad k = 1, \ldots, K , \qquad (7)$$

with $P_{k,\max}$ the maximum allowed transmit power for the $k$-th user, the following result can be proven.

**Proposition 1:** *The non-cooperative game defined in (7) admits a unique NE point $(p_k^*)$, for $k = 1, \ldots, K$, wherein $p_k^* = \min\{\bar{p}_k, P_{k,\max}\}$, with $\bar{p}_k$ the $k$-th user transmit power such that the $k$-th user maximum SINR $\gamma_k^*$ equals $\bar{\gamma}$, i.e. the unique solution of the equation $f(\gamma) = \gamma f'(\gamma)$, with $f'(\gamma)$ the derivative of $f(\gamma)$.*

---

[1] To simplify subsequent notation, we assume that the transmitted power $p_k$ subsumes also the gain of the transmit and receive antennas.

[2] We assume here, for simplicity, a real channel model; generalization to practical channels, with I and Q components, is straightforward.

**Proof:** The proof is here sketched as a lead-in to the exposition of the full MIMO case of the forthcoming Section 3.3. According to theorem 11 in [3], a NE in a non-cooperative game exists if the strategy set $\mathscr{S}_k$ is a nonempty, convex, and compact subset of an Euclidean space, and if the utility function of each player of the game is quasi-concave in its own power (this means that there exists a point below which the function is non-decreasing, and above which the function is non-increasing). In the considered game, we have $\mathscr{S}_k = [0, P_{k,\max}]$, so the former condition is obviously fulfilled; to verify the latter condition, it suffices to show that the utility function $u_k$ is increasing in an $\varepsilon$-neighborhood of $p_k = 0$ and that the first order partial derivative of $u_k$ with respect to $p_k$ has only one zero for $p_k > 0$. Note that the utility $u_k$ equals zero for $p_k = 0$, and is positive for $p_k = 0^+$, thus implying that it is an increasing function for $p_k \in [0, \varepsilon]$. Consider now the partial derivative of $u_k(\cdot)$ with respect to $p_k$ and equate it to zero; since, given Eq. (6), it is seen that $\gamma_k = p_k d\gamma_k/dp_k$, each user's utility is maximized if each user is able to achieve the SINR $\bar{\gamma}$, that is the unique[3] solution of the equation $f(\gamma) = \gamma f'(\gamma)$. The existence of an NE is thus proven. Given the uniqueness of the utility-maximizing SINR $\bar{\gamma}$, and the bi-injective correspondence between the achieved SINR and the transmit power for each user, the above NE is also unique. ∎

In practice, the above NE is reached through the following iterative algorithm. Given any set of transmit powers, the standard power control iterations as detailed in [10] are used so that each user may either achieve its target SINR or, should this be not possible, transmit at its maximum allowed power.

### 3.2 Transmit power control and choice of the linear receiver

Consider now the following non-cooperative game:

$$\max_{p_k \in [0, P_{k,\max}], \mathbf{d}_k \in \mathscr{R}^{N_R}} \frac{f(\gamma_k)}{p_k}, \quad k = 1, \ldots, K. \quad (8)$$

We have now

$$\max_{p_k, \mathbf{d}_k} \frac{f(\gamma_k)}{p_k} = \max_{p_k} \frac{f(\max_{\mathbf{d}_k} \gamma_k)}{p_k}, \quad (9)$$

i.e. we can first take care of SINR maximization with respect to $\mathbf{d}_k$ and then consider the problem of utility maximization with respect to $p_k$. It is well-known that, among linear receivers, the minimum mean square error (MMSE) receiver is the one that maximizes the SINR. As a consequence, we have the following result.

**Proposition 2:** *The non-cooperative game defined in (8) admits a unique[4] NE point $(p_k^*, \mathbf{d}_k^*)$, for $k = 1, \ldots, K$, wherein*

- $\mathbf{d}_k^* = \sqrt{p_k} \mathbf{M}^{-1} \mathbf{H}_k \mathbf{a}_k$ *is the MMSE receiver for the k-th user, with $M = (\sum_{k=1}^{K} p_k \mathbf{H}_k \mathbf{a}_k \mathbf{a}_k^T \mathbf{H}_k^T + \frac{\mathcal{N}_0}{2} \mathbf{I}_{N_R})$ the data covariance matrix. Denote by $\gamma_k^*$ the corresponding SINR.*
- $p_k^* = \min\{\bar{p}_k, P_{k,\max}\}$, *with $\bar{p}_k$ the k-th user transmit power such that the k-th user maximum SINR $\gamma_k^*$ equals*

---
[3]Uniqueness of $\bar{\gamma}$ is ensured by the fact that the efficiency function is S-shaped [3].
[4]Here and in the following uniqueness with respect to the receiver $\mathbf{d}_k$ is meant up to a positive scaling factor.

*$\bar{\gamma}$, i.e. the unique solution of the equation $f(\gamma) = \gamma f'(\gamma)$, with $f'(\gamma)$ the derivative of $f(\gamma)$.*

**Proof:** For the sake of brevity, we just sketch the key parts of this proof. We have already discussed the fact that SINR maximization requires that the receiver $\mathbf{d}_k$ is the MMSE detector. It is easy to show that, with MMSE detection, the SINR for the $k$-th user is expressed as $\gamma_k = p_k \mathbf{a}_k^T \mathbf{H}_k^T \mathbf{M}_k^{-1} \mathbf{H}_k \mathbf{a}_k$, with $\mathbf{M}_k = \mathbf{M} - p_k \mathbf{H}_k \mathbf{a}_k \mathbf{a}_k^T \mathbf{H}_k^T$ the covariance matrix of the interference seen by the $k$-th user. Note that the relation $\gamma_k = p_k d\gamma_k/dp_k$ still holds here, thus implying that the arguments of the proof of Proposition 1 can be easily borrowed in order to show existence and uniqueness of the NE for the game (8). ∎

In practice, the above NE is reached through the following iterative algorithm. Given any set of transmit powers, each user sets its uplink receiver equal to the MMSE receiver. After that, users adjust their transmit power in order to achieve the target SINR, using the standard power control iterations of [10]. These steps are repeated until convergence is reached.

### 3.3 Transmit power control, beamforming, and choice of the linear receiver

Finally, let us now consider the more challenging case in which utility maximization is performed with respect to the transmit power, beamforming vector and choice of the uplink linear receiver, i.e.

$$\max_{p_k \in [0, P_{k,\max}], \mathbf{d}_k \in \mathscr{R}^{N_R}, \mathbf{a}_k \in \mathscr{R}_1^{N_T}} \frac{f(\gamma_k)}{p_k}, \quad k = 1, \ldots, K, \quad (10)$$

with $\mathscr{R}_1^{N_T}$ the set of unit-norm $N_T$-dimensional vectors with real entries. Note that

$$\max_{p_k, \mathbf{d}_k, \mathbf{a}_k} \frac{f(\gamma_k)}{p_k} = \max_{p_k} \frac{f(\max_{\mathbf{d}_k, \mathbf{a}_k} \gamma_k)}{p_k}. \quad (11)$$

Given the above equation, we have to consider first the problem of SINR maximization with respect to the vectors $\mathbf{d}_k$ and $\mathbf{a}_k$. Again, the SINR-maximizing linear receiver is the MMSE receiver. Since, as already discussed, the $k$-th user SINR for MMSE detection is $\gamma_k = p_k \mathbf{a}_k^T \mathbf{H}_k^T \mathbf{M}_k^{-1} \mathbf{H}_k \mathbf{a}_k$, it is easily seen that the SINR-maximizing beamforming vector $\mathbf{a}_k$ is the eigenvector corresponding to the maximum eigenvalue of the matrix $\mathbf{H}_k^T \mathbf{M}_k^{-1} \mathbf{H}_k$. Of course, the question now arises if, when such beamforming vector update is cyclically performed by the active users, a stable equilibrium is reached. The following result holds.

**Theorem:** *Assume that the active users cyclically update their beamforming vectors in order to maximize their own achieved SINR at the output of a linear MMSE receiver. This procedure converges to a fixed point.*

**Proof:** Denote by $\mathbf{a}_1, \ldots, \mathbf{a}_K$ the set of current beamformers for the active users. The system sum capacity is well-known to be expressed as

$$\mathscr{C}_{\text{SUM}} = \tfrac{1}{2} \log(\det(\mathbf{M})) - \tfrac{1}{2} \log\left(\det\left(\tfrac{\mathcal{N}_0}{2} \mathbf{I}_{N_R}\right)\right) =$$
$$\tfrac{1}{2} \log(\det(\mathbf{M}_k + p_k \mathbf{H}_k \mathbf{a}_k \mathbf{a}_k^T \mathbf{H}_k^T)) - \tfrac{1}{2} \log\left(\det\left(\tfrac{\mathcal{N}_0}{2} \mathbf{I}_{N_R}\right)\right). \quad (12)$$

Exploiting the relation $\det(\mathbf{A} + \mathbf{x}\mathbf{y}^T) = \det(\mathbf{A})(1 + \mathbf{y}^T\mathbf{A}^{-1}\mathbf{x})$, the sum capacity is also written as

$$\mathscr{C}_{\text{SUM}} = \frac{1}{2}\log\left(\det(\mathbf{M}_k)\left(\frac{2}{\mathscr{N}_0}\right)^{N_R}\right) + \frac{1}{2}\log\left(1 + \underline{p_k \mathbf{a}_k^T \mathbf{H}_k^T \mathbf{M}_k^{-1} \mathbf{H}_k \mathbf{a}_k}\right). \quad (13)$$

The underlined term in the above equation is the *k*-th user SINR at the output of its MMSE receiver. Accordingly, if the *k*-th user updates its beamforming vector with the eigenvector corresponding to the maximum eigenvalue of the matrix $\mathbf{H}_k^T \mathbf{M}_k^{-1} \mathbf{H}_k$, the system sum capacity is increased. Iterating this reasoning, it can be shown that every time that a user updates its own beamforming vector this leads to an increase of the system sum capacity. Since sum capacity is obviously upper bounded, this procedure must admit a fixed point. ∎

Equipped with the above result, and assuming[5] that the equilibrium SINR resulting for the non-cooperative SINR maximization game with respect to the vectors $\mathbf{d}_k$ and $\mathbf{a}_k$ is continuous with respect to the transmit powers $p_1, p_2, \ldots, p_K$, we are now ready to state our result on the game (10).

**Proposition 3:** *The non-cooperative game defined in (10) admits a NE point $(p_k^*, \mathbf{d}_k^*, \mathbf{a}_k^*)$, for $k = 1, \ldots, K$, wherein*

- $\mathbf{a}_k^*$ *and $\mathbf{d}_k^*$ are the equilibrium k-th user beamforming vector and receive filter resulting from the non-cooperative SINR maximization game. Denote by $\gamma_k^*$ the corresponding SINR.*
- $p_k^* = \min\{\bar{p}_k, P_{k,\max}\}$, *with $\bar{p}_k$ the k-th user transmit power such that the k-th user maximum SINR $\gamma_k^*$ equals $\bar{\gamma}$, i.e. the unique solution of the equation $f(\gamma) = \gamma f'(\gamma)$, with $f'(\gamma)$ the derivative of $f(\gamma)$.*

**Proof:** The complete proof is omitted due to lack of space. Note that the considered game can be seen as the composition of two separable games, namely the power control game and the beamformer plus receive filter game. The former game admits a unique NE based on Proposition 1, while the latter game admits a NE based on the previous theorem on sum capacity. Exploiting the results of [11], the existence of a NE for the two subgames implies the existence of a NE for the game (10). ∎

## 4. NON-COOPERATIVE RESOURCE ALLOCATION: NON-LINEAR SIC RECEIVER

Consider now the case in which a non-linear decision feedback receiver is used at the receiver. We assume that the users are indexed according to a non-increasing sorting of their channel-induced signatures, i.e. we assume that $\|\mathbf{H}_1\mathbf{a}_1\| > \|\mathbf{H}_2\mathbf{a}_2\| > \ldots, \|\mathbf{H}_K\mathbf{a}_K\|$. We consider a SIC receiver wherein detection of the bit from the *k*-th user is made according to the following rule

$$\widehat{b}_k = \text{sign}\left[\mathbf{d}_k^T\left(\mathbf{r} - \sum_{j<k}\sqrt{p_j}\mathbf{H}_j\mathbf{a}_j\widehat{b}_j\right)\right]. \quad (14)$$

Otherwise stated, when detecting a certain symbol, the contribution from the information symbols that have been already detected is subtracted from the received data. If past decisions are correct, users that are detected later enjoy a considerable reduction of multiple access interference, and indeed the SINR for user *k*, under the assumption of correctness of past decisions, is written as

$$\gamma_k = \frac{p_k(\mathbf{d}_k^T\mathbf{H}_k\mathbf{a}_k)^2}{\frac{\mathscr{N}_0}{2}\|\mathbf{d}_k\|^2 + \sum_{j>k} p_j(\mathbf{d}_k^T\mathbf{H}_j\mathbf{a}_j)^2}. \quad (15)$$

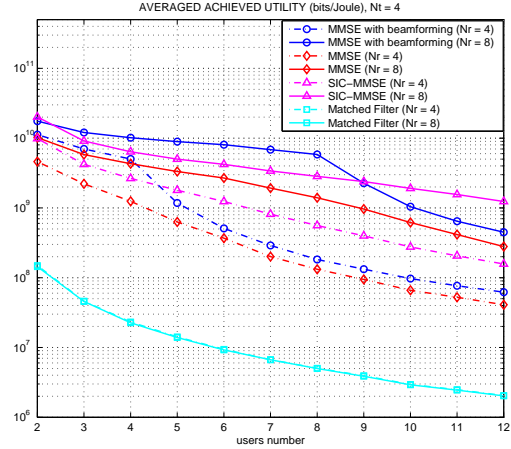

Figure 1: Achieved average utility at the NE versus the users' number for the proposed non-cooperative games.

Now, given receiver (14) and the SINR expression (15), we consider here the problem of utility maximization with respect to the transmit power, and to the choice of the receivers $\mathbf{d}_1, \ldots, \mathbf{d}_K$, i.e.:

$$\max_{p_k, \mathbf{d}_k} \frac{f(\gamma_k(p_k, \mathbf{d}_k))}{p_k}, \quad \forall k = 1, \ldots, K. \quad (16)$$

The following result can be shown to hold.

**Proposition 4:** *Define $\tilde{\mathbf{H}}_k = [\mathbf{H}_k\mathbf{a}_k, \ldots, \mathbf{H}_K\mathbf{a}_K]$, and $\mathbf{P}_k = \text{diag}(p_k, \ldots, p_K)$. The non-cooperative game defined in (16) admits a unique NE point $(p_k^*, \mathbf{d}_k^*)$, for $k = 1, \ldots, K$, wherein*

- $\mathbf{d}_k^* = \sqrt{p_k}(\tilde{\mathbf{H}}_k\mathbf{P}_k\tilde{\mathbf{H}}_k^T + \frac{\mathscr{N}_0}{2}\mathbf{I}_N)^{-1}\mathbf{H}_k\mathbf{a}_k$ *is the unique k-th user receive filter[6] that maximizes the k-th user's SINR $\gamma_k$ given in (15). Denote $\gamma_k^* = \max_{\mathbf{d}_k} \gamma_k$.*
- $p_k^* = \min\{\bar{p}_k, P_{k,\max}\}$, *with $\bar{p}_k$ the k-th user transmit power such that the k-th user maximum SINR $\gamma_k^*$ equals $\bar{\gamma}$, i.e. the unique solution of the equation $f(\gamma) = \gamma f'(\gamma)$, with $f'(\gamma)$ the derivative of $f(\gamma)$.*

**Proof:** The proof is omitted here due to lack of space. It can be made along the same track that led to the proof of Proposition 2. ∎

## 5. NUMERICAL RESULTS

In this section we present some simulation results that give insight into the performance of the proposed non-cooperative resource allocation policies.

---
[5]Actually, this assumption has been numerically tested; its formal proof however, appears a little involved and is the object of current investigation.

[6]Uniqueness is here up to a positive scaling factor.

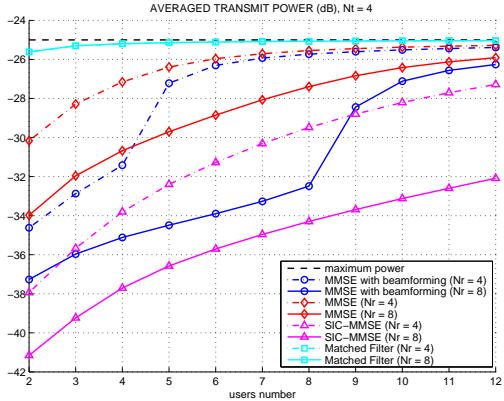

Figure 2: Average transmit power at the NE versus the users' number for the proposed non-cooperative games.

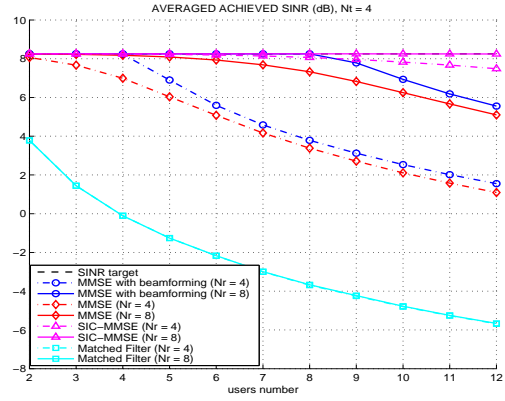

Figure 3: Achieved average SINR at the NE versus the users' number for the proposed non-cooperative games.

We consider an uplink multiuser MIMO system using uncoded BPSK and consider the corresponding efficiency function $f(\gamma_k) = (1 - e^{-\gamma_k})^M$. We consider $N_T = 4$ transmit antennas for each user, and assume that the packet length is $M = 120$; for this value of $M$ the equation $f(\gamma) = \gamma f'(\gamma)$ can be shown to admit the solution $\bar{\gamma} = 6.689 = 8.25$dB. The system data rate is $R = 10^5$bps. A single-cell system is considered, wherein users may have random positions with a distance from the AP ranging from 10m to 1000m. The channel matrix $\mathbf{H}_k$ for the generic $k$-th user is assumed to have Rayleigh distributed entries with mean equal to $d_k^{-1}$, with $d_k$ being the distance of user $k$ from the AP. We take the ambient noise level to be $\mathcal{N}_0 = 10^{-9}$W/Hz, while the maximum allowed power $P_{k,\max}$ is $-25$dBW. We present the results of averaging over 3000 independent realizations for the users locations and fading channel coefficients. The beamforming vector of the generic $k$-th user is chosen as the eigenvector corresponding to the maximum eigenvalue of the matrix $\mathbf{H}_k^T \mathbf{H}_k$; this vector is then used as the starting point for the games that include beamformer optimization, and as the (constant) beamformer for the remaining games.

Figs. 1 - 3 show the achieved average utility (measured in bits/Joule), the average user transmit power and the average achieved SINR at the receiver output versus the number of users, for the several considered games, and for a $4 \times 4$ and $4 \times 8$ MIMO system. Inspecting the curves, it is seen that smart resource allocation algorithms may bring very remarkable performance improvements. As an example, for $K = 10$ users and $N_R = 8$ the utility achieved by the SIC-MMSE game and by the MMSE+Beamforming game is about 660 and 330 times larger (!!) than the utility achieved by the power allocation game coupled with a matched filter, respectively. Interestingly, it is seen that the SIC-MMSE game outperforms the MMSE+Beamforming game for low number of users and for large number of users: indeed, in a lightly loaded system beamforming may not yield substantial performance improvements, while, for heavily loaded systems, SIC processing is extremely beneficial. It is also seen from Fig. 3 that in many instances receivers achieve on the average an output SINR that is smaller than the target SINR $\bar{\gamma}$: indeed, due to fading and distance path losses, achieving the target SINR would require some users to a transmit at higher power than the maximum allowed power $P_{k,\max}$, and so these users are not able to achieve the optimal target SINR. Of course, the use of cross-layer resource allocation procedures help reducing the gap between the average and target SINRs. Finally, note that results confirm that increasing the number of receive antennas improves the system performance.

Overall, it can be stated that the use of cross-layer resource allocation policies brings very significant performance improvements to the energy efficiency of a multiuser MIMO system.